\documentclass[preprint,prd,showpacs,amsmath,amssymb,amsthm,nofootinbib]{revtex4}
\usepackage[mathscr]{euscript}
\usepackage{bm}
\usepackage{graphicx}
\usepackage{subfigure}
\usepackage{overpic}
\usepackage{multirow}
\usepackage{booktabs}
\usepackage{array}
\usepackage{floatrow}
\usepackage{float} 
\usepackage{amsmath,amssymb,amsthm}
\usepackage{cases}
\usepackage[colorlinks=true,linkcolor=red]{hyperref}
\newfloatcommand{capbtabbox}{table}[][\FBwidth]
\newcommand{\bea}{\begin{eqnarray}}
\newcommand{\eea}{\end{eqnarray}}
\newcommand{\beq}{\begin{equation}}
\newcommand{\eeq}{\end{equation}}

\def\/{\over}

\begin{document}
	
\title{Frame-Dependence of the Hamilton-Jacobi Formalism for Inflation and Reheating in Non-Minimal Gravity}

\author{ Feng-Yi Zhang}
\email{zfy@usc.edu.cn}
\affiliation{School of Mathematics and Physics, University of South China, Hengyang, 421001, China}
\affiliation{Hunan Key Laboratory of Mathematical Modeling and Scientific Computing, University of South China, Hengyang, 421001, China}

\author{Li-Yang Chen}
\affiliation{College of Physics and Engineering Technology, Chengdu Normal University, Chengdu, Sichuan 611130, China}
\author{ Rongrong Zhai}
\email{Corresponding author: rrzhai@xztu.edu.cn}
\affiliation{Department of Physics, Xinzhou Normal University, Xinzhou 034000, Shanxi, China}


\begin{abstract}
In this work, we investigate the Hamilton-Jacobi formalism for non-minimally coupled inflation, focusing on the methodological frame-dependence arising from its application in the Jordan and Einstein frames. We systematically compare the physical predictions from two distinct computational schemes: applying the Hamilton-Jacobi approximation before versus after the conformal transformation. This comparison is conducted for both the metric and Palatini formalisms. Our results, consistent with Planck  data, reveal significant quantitative differences between the two schemes, highlighting a subtle frame-dependence in the approximation method. These discrepancies, observed in the spectral index, the tensor-to-scalar ratio, and reheating parameters, are more pronounced in the Palatini formalism. Our study emphasizes the sensitivity of cosmological predictions to the computational path chosen, and provides a quantitative analysis of this methodological uncertainty, offering valuable insights into the robustness of predictions in modified gravity.
\end{abstract}

\maketitle

\section{Introduction}\label{sec1}
Inflation describes a period of accelerated expansion in the early universe, resolving key issues in the standard cosmological model, such as the flatness, horizon, and monopole problems~\cite{STAROBINSKY198099, PhysRevLett.48.1220, PhysRevD.23.347, LINDE1982389}. It also predicts a nearly scale-invariant spectrum of primordial perturbations, consistent with current Cosmic Microwave Background (CMB) observations. While large-scale fluctuations are well-constrained, small-scale perturbations remain poorly understood due to observational limitations, leaving open questions regarding primordial black holes and stochastic gravitational waves~\cite{10.1093/mnras/152.1.75, 10.1093/mnras/168.2.399, PhysRevD.96.063503, Di_2018, PhysRevD.100.023537, PhysRevD.101.023505, PhysRevD.106.023517, PhysRevD.106.063537}. As a result, inflationary models are primarily tested through indirect evidence from CMB data.

CMB observations provide tight constraints on inflationary observables such as the spectral index and the tensor-to-scalar ratio, leading to the exclusion of certain simple inflationary models with monomial potentials~\cite{akrami2020planck, PhysRevLett.127.151301}. However, when considering extensions to General Relativity (GR), such as non-minimally coupled scalar-tensor theories, these models can reconcile with the data~\cite{refId0, PhysRevD.88.023529, PhysRevLett.105.011302}. The distinction between the metric and Palatini formalisms is crucial in such theories, as the gravitational degrees of freedom and resulting dynamics can differ significantly~\cite{BAUER2008222, PhysRevD.83.044018, Tenkanen_2017, PhysRevD.97.083513, Carrilho_2018, Almeida_2019, PhysRevD.101.063517, PhysRevD.101.084007, FAN2015230, PhysRevD.96.103542}. Notably, in the Palatini formalism, the tensor-to-scalar ratio tends to be suppressed~\cite{R_s_nen_2017}.

Following inflation, the universe enters the reheating phase, during which the inflaton decays into radiation, setting the initial conditions for Big Bang nucleosynthesis (BBN)~\cite{DOLGOV1982329, ABBOTT198229, PhysRevLett.48.1437,PhysRevD.42.2491, PhysRevD.51.5438, Amin_2014, RevModPhys.78.537, PhysRevD.56.6484, PhysRevD.56.3258, PhysRevLett.73.3195, PhysRevD.56.6175}. While direct observations of reheating remain scarce, indirect constraints can be derived by linking inflationary parameters to post-inflationary observables through the evolution of the Hubble scale~\cite{PhysRevD.82.023511,  PhysRevLett.114.081303, Cook_2015, PhysRevD.92.063506, PhysRevD.103.103540, PhysRevLett.113.041302, PhysRevD.104.103530, PhysRevD.93.083524, PhysRevD.95.103502, Goswami_2018, Maity_2019, DENG2022101135, Mishra_2021, Gong_2015, PhysRevD.102.021301, ZHANG2023101169, ZHANG2024101482, ZHANG2024138765}. This connection allows us to estimate reheating parameters, such as the reheating temperature and the number of e-folds, which provides important consistency checks for inflationary models.

Recently, the Hamilton-Jacobi formalism offers an alternative approach to inflationary dynamics by expressing the Hubble parameter as a function of the scalar field rather than directly specifying the potential~\cite{PhysRevD.42.3936, 2000Cosmological, Muslimov_1990, LIDSEY199142, PhysRevD.50.7222, PhysRevD.56.2002, PhysRevD.68.043508, PhysRevD.90.084028, Villanueva_20152, Sheikhahmadi_2016, PhysRevD.95.023501, Videla_2017, ASADI2019114827, Gabriel2020, PhysRevLett.122.191301, PhysRevD.102.103517, Yang_2024, Cicciarella_2019}. This approach is particularly useful in modified gravity theories, where it facilitates the reconstruction of inflationary potentials and helps avoid certain perturbative issues~\cite{Sheikhahmadi_2016}. Although the Hamilton-Jacobi method has been applied to non-minimally coupled inflation~\cite{PhysRevD.72.043523, Zhang:2025tpg}, which focus on a single frame or gravitational formalism. Additionally, the application of the Hamilton-Jacobi method to study the reheating phase remains relatively scarce.

In this study, we conduct a systematic comparison of the Hamilton-Jacobi formalism as applied in the Jordan and Einstein frames for non-minimally coupled inflation, considering both metric and Palatini formalisms. Our goal is to quantify how the choice of frame for implementing the Hamilton-Jacobi approximation scheme affects the final predictions for both inflationary and reheating observables. By starting with a prescribed Hubble parameter, we derive effective potentials corresponding to each computational path and compute the resulting physical predictions. This analysis provides a unified perspective on the methodological frame-dependence of cosmological predictions, offering new insights into the robustness of theoretical calculations in modified gravity.

This paper is organized as follows. Section~\ref{sec2} introduces the background equations of non-minimally coupled gravity in both the Jordan and Einstein frames, under the metric and Palatini formulations. Section~\ref{sec3} develops the Hamilton-Jacobi method in both frames to study inflationary dynamics. In Section~\ref{sec4}, we analyze the reheating phase and establish its connection to inflationary observables. Section~\ref{sec5} presents numerical results and discusses frame- and formulation-dependent differences. We conclude in Section~\ref{sec6}.
Throughout, we adopt the metric signature $(-,+,+,+)$, set $c = \hbar = 1$, and use the reduced Planck mass $M_{\mathrm{pl}} = 1/\sqrt{8\pi G} = 1$.


\section{Background Equations for Non-Minimally Coupled Inflationary Models}
\label{sec2}

In this section, we outline the background cosmological equations in non-minimally coupled inflationary models, focusing on both the metric and Palatini formalisms. The analysis is carried out in both the Jordan and Einstein frames, highlighting how the formulation and conformal frame affect the dynamical evolution of the universe.

\subsection{Jordan Frame Formulation}

In the Jordan frame, the scalar field $\phi$ is non-minimally coupled to gravity via a function of the Ricci scalar. The action takes the form
\begin{align}
	S_J = \int d^4x \sqrt{-g} \left[ \frac{1}{2}f(\phi)R(\Gamma) - \frac{1}{2}g^{\mu\nu}\partial_\mu\phi \partial_\nu\phi - V(\phi) \right],
	\label{ac}
\end{align}
where $g$ is the determinant of the metric $g_{\mu\nu}$, $V(\phi)$ is the scalar potential, and $f(\phi) \equiv 1 - \xi \phi^2$ is the non-minimal coupling function~\cite{BEZRUKOV2008703}. The coupling constant $\xi$ determines the strength of the interaction between the scalar field and the Ricci scalar, with $\xi = 0$ recovering minimal coupling.
The Ricci scalar $R(\Gamma)$ depends on the choice of gravitational formalism. In the metric formulation, the affine connection $\Gamma$ is the Levi-Civita connection derived from the metric: $\Gamma = \tilde{\Gamma}(g_{\mu\nu})$. In contrast, the Palatini formulation treats $\Gamma$ as an independent variable. In this case, the connection is modified according to~\cite{BAUER2008222}:
\begin{align}
    \Gamma^{\gamma}_{\mu\nu} = \tilde{\Gamma}^{\gamma}_{\mu\nu} + \delta^\gamma_\mu \partial_\nu \ln \sqrt{f(\phi)} + \delta^\gamma_\nu \partial_\mu \ln \sqrt{f(\phi)} - g_{\mu\nu} \partial^\gamma \ln \sqrt{f(\phi)}.
    \label{GA}
\end{align}
When $f(\phi) = 1$, i.e., $\xi = 0$, both formalisms reduce to standard GR, yielding identical field equations. However, once non-minimal coupling is introduced, the equations differ significantly, leading to distinct cosmological predictions.

Assuming a spatially flat Friedmann–Robertson–Walker (FRW) metric,
\begin{align}
	ds^2 = -dt^2 + a^2(t)\delta_{ij}dx^idx^j,
	\label{FRW}
\end{align}
the background equations for the non-minimally coupled scalar field can be derived. Introducing $\alpha \equiv \xi \phi^2$ and $\beta \equiv 1 - \alpha\left[1 - (1 - \sigma)6\xi\right]$, we obtain:
\begin{align}
	3H^2 &= \frac{1}{1 - \alpha} \left[ \frac{1}{2} \dot{\phi}^2 + V(\phi) + 6\xi H\phi\dot{\phi} - \frac{3\sigma\xi\alpha\dot{\phi}^2}{1 - \alpha} \right],
	\label{F1} \\
	\ddot{\phi} + 3H\dot{\phi} &+ \frac{(1 - \alpha)V_{,\phi}(\phi)}{\beta} + \frac{\xi\phi \left[ 4V(\phi) - \left(1 - (1 - \sigma)6\xi\right)\dot{\phi}^2 \right]}{\beta} = 0,
	\label{p}
\end{align}
where $H = \dot{a}/a$ is the Hubble parameter, the dot denotes a derivative with respect to cosmic time $t$, and the parameter $\sigma$ distinguishes between the metric ($\sigma = 0$) and Palatini ($\sigma = 1$) formulations.

\subsection{Einstein Frame Formulation}
To simplify the gravitational action to the standard Einstein-Hilbert form and work with a canonically normalized kinetic term for the scalar field, it is conventional to perform a conformal transformation from the Jordan frame to the Einstein frame \cite{Sotiriou:2008rp, Baumann:2009ds}. This transformation is a mathematical tool that maps the theory to a frame where the gravitational dynamics are simpler to analyze, though the physical predictions for observable quantities, when calculated exactly, must be frame-independent.
	
A crucial point in non-minimally coupled theories is the distinction between the metric and Palatini formalisms. In the metric formalism, the connection is assumed a priori to be the Levi-Civita connection of the metric $g_{\mu\nu}$. In the Palatini formalism, the metric and the connection are treated as independent variables, and the connection is solved for from its own equation of motion. While these formalisms are equivalent in General Relativity, they lead to different physical predictions in modified gravity theories, including the non-minimally coupled model studied here~\cite{BAUER2008222}. As shown below, this distinction is encoded in the parameter $\sigma$, which affects the field redefinition and subsequent dynamics.
To simplify the equations and analyze the inflationary dynamics from a geometric perspective, we perform a conformal transformation to the Einstein frame:
\begin{align}
	\hat{g}_{\mu\nu} = f(\phi) g_{\mu\nu},
	\label{tran}
\end{align}
where quantities in the Einstein frame are denoted by a hat. The transformed line element becomes:
\begin{align}
	d\hat{s}^2 = f(\phi) ds^2 = -d\hat{t}^2 + \hat{a}^2(\hat{t})\delta_{ij}dx^idx^j,
	\label{ds}
\end{align}
leading to the relations
\begin{align}
	\hat{a}(\hat{t}) = \sqrt{f(\phi)} a(t), \quad d\hat{t} = \sqrt{f(\phi)} dt.
	\label{at}
\end{align}
Applying this transformation to the action~\eqref{ac}, we obtain the Einstein frame action:
\begin{align}
	S_E = \int d^4x \sqrt{-\hat{g}} \left[ \frac{1}{2} \hat{R} - \frac{1}{2} \hat{g}^{\mu\nu} \partial_\mu \hat{\phi} \partial_\nu \hat{\phi} - \hat{V}(\hat{\phi}) \right],
	\label{ace}
\end{align}
where the canonical scalar field $\hat{\phi}$ and the potential $\hat{V}(\hat{\phi})$ are given by:
\begin{align}
	\frac{d\hat{\phi}}{d\phi} = \frac{\sqrt{\beta}}{1 - \alpha}\label{newp}, \\
	\hat{V}(\hat{\phi}) = \frac{V(\phi)}{f^2(\phi)}.
    \label{pots}
\end{align}

In the Einstein frame, the background equations take the standard GR form:
\begin{align}
	&3\hat{H}^2 = \frac{1}{2} \hat{\phi}^{\prime 2} + \hat{V}(\hat{\phi}), \label{FE1} \\
	&\hat{\phi}^{\prime\prime} + 3\hat{H} \hat{\phi}^\prime + \hat{V}_{,\hat{\phi}}(\hat{\phi}) = 0,
	\label{chi}
\end{align}
where primes denote derivatives with respect to the rescaled time $\hat{t}$.
The Hubble parameter and scalar field velocity in the Einstein frame are related to their Jordan frame counterparts by:
\begin{align}
	\hat{\phi}^\prime &= \frac{d\hat{\phi}}{d\phi} \frac{d\phi}{dt} \frac{dt}{d\hat{t}} 
	= \frac{\sqrt{\beta}}{(1 - \alpha)^{3/2}} \dot{\phi}, \label{pd} \\
	\hat{H} &= \frac{1}{\sqrt{f}} \left( H + \frac{\dot{f}}{2f} \right).
	\label{HE}
\end{align}
This formalism provides a useful foundation for the Hamilton–Jacobi analysis, which will be conducted in both frames in the subsequent sections.

\section{Hamilton-Jacobi Approach and Inflationary Dynamics}
\label{sec3}

\subsection{Hamilton–Jacobi Method in the Jordan Frame}

We investigate the dynamics of non-minimally coupled inflation using the Hamilton-Jacobi formalism~\cite{PhysRevD.42.3936}, applied to both the metric and Palatini formulations. 
To ensure a quasi-exponential expansion during inflation, the standard slow-roll conditions must be satisfied:
$|\dot{\phi}/\phi|\ll H$, $|\ddot{\phi}/\dot{\phi}|\ll H$, and $\dot{\phi}^2\ll V(\phi)$. 
Under these conditions, the background equations~\eqref{F1} and~\eqref{p} simplify to
\begin{align}
	&3(1 - \alpha) H^2 \simeq V(\phi), \label{F11} \\
	&3H\dot{\phi} \simeq -\frac{4\xi\phi V(\phi) + (1 - \alpha)V_{,\phi}(\phi)}{\beta}. \label{p1}
\end{align}
To solve these equations, we consider the Hubble parameter as a function of the inflaton field, $H = H(\phi)$.
By differentiating Eq.~\eqref{F11} with respect to $\phi$ and substituting the result into Eq.~\eqref{p1}, we derive 
\begin{align}
	\dot{\phi} \simeq -\frac{2(1 - \alpha) \mathcal{H}(\phi)}{\beta}, \label{dp}
\end{align}
where $\mathcal{H}(\phi) \equiv \xi\phi H(\phi) + (1 - \alpha) H_{,\phi}(\phi)$.
Substituting Eq.~\eqref{dp} into the Friedmann equation~\eqref{F1}, we derive the Hamilton-Jacobi equation:
\begin{align}
	3(1 - \alpha) H^2(\phi) &= V(\phi) + \frac{2(1 - \alpha)^2 \mathcal{H}^2(\phi)}{\beta^2} 
	- \frac{12\xi\phi(1 - \alpha) H(\phi)\mathcal{H}(\phi)}{\beta} \nonumber \\
	&\quad - \frac{12\xi\sigma\alpha(1 - \alpha)\mathcal{H}^2(\phi)}{\beta^2}. \label{H}
\end{align}
Solving for the potential yields:
\begin{align}
	V(\phi) &= 3(1 - \alpha) H^2(\phi) - \frac{2(1 - \alpha)^2 \mathcal{H}^2(\phi)}{\beta^2} 
	+ \frac{12\xi\phi(1 - \alpha) H(\phi)\mathcal{H}(\phi)}{\beta} \nonumber \\
	&\quad + \frac{12\xi\sigma\alpha(1 - \alpha)\mathcal{H}^2(\phi)}{\beta^2}. \label{P}
\end{align}

\subsection{Hamilton–Jacobi Method in the Einstein Frame}

Applying the slow-roll approximation to the Einstein frame Eqs.~\eqref{FE1} and~\eqref{chi} leads to
\begin{align}
	3\hat{H}^2 &\simeq \hat{V}(\hat{\phi}), \label{FE2} \\
	3\hat{H} \hat{\phi}' &\simeq -\hat{V}_{,\hat{\phi}}(\hat{\phi}). \label{chi1}
\end{align}
From this, we obtain
\begin{align}
\hat{\phi}' \simeq -2\hat{H}_{,\hat{\phi}}.
\end{align}
Substituting this into the Friedman Eq. \eqref{FE1} we obtain the Hamilton-Jacobi equation in the Einstein frame:
\begin{align}
	3\hat{H}^2(\hat{\phi}) = \hat{V}(\hat{\phi}) + 2\hat{H}_{,\hat{\phi}}^2~ , \label{Hamilton-JacobiE}
\end{align}
from which the potential is derived as
\begin{align}
	\hat{V}(\hat{\phi}) = 3\hat{H}^2(\hat{\phi}) - 2\hat{H}_{,\hat{\phi}}^2~. \label{VE}
\end{align}

\subsection{Comparing Methodological Frames for Inflationary Observables}
In this section, we compare the application of the Hamilton-Jacobi method in the Jordan and Einstein frames. It is crucial to clarify a key methodological point. Fundamentally, the Einstein frame potential $\hat{V}(\hat{\phi})$ is uniquely determined by the Jordan frame potential $V(\phi)$ via the conformal transformation~\eqref{pots}.   
    However, the Hamilton-Jacobi formalism, which relies on the slow-roll approximation, can be applied in two distinct computational paths: 
	\begin{itemize}
		\item[(i)] applying the slow-roll approximation in the Jordan frame first to find an effective potential $V(\phi)$, and then transforming this result to the Einstein frame.
		\item[(ii)] transforming the kinematic variables (e.g., the Hubble parameter) to the Einstein frame first, and then applying the slow-roll approximation directly within the Einstein frame.
	\end{itemize}
    These two procedures are not mathematically identical and lead to slightly different effective potentials, which allows for a quantitative comparison of how the choice of frame for the approximation scheme impacts the final observables.
    
To facilitate this comparison, we define two effective potentials that arise from these two distinct computational paths. First, we take the potential $V(\phi)$ derived from the Hamilton-Jacobi method in the Jordan frame~\eqref{P} and transform it into the Einstein frame using the exact relation~\eqref{pots}. We denote this potential by $\hat{V}_J$ to signify its origin in the Jordan-frame approximation scheme:
	\begin{align}
		\hat{V}_J \equiv \frac{V(\phi)}{(1 - \alpha)^2} &= \frac{3H^2(\phi)}{1 - \alpha} - \frac{2\mathcal{H}^2(\phi)}{\beta^2}
		+ \frac{12\xi\phi H(\phi)\mathcal{H}(\phi)}{\beta(1 - \alpha)} + \frac{12\xi\sigma\alpha \mathcal{H}^2(\phi)}{\beta^2(1 - \alpha)}, \label{VJ}
	\end{align}
where the subscript ``J" denotes the computational path originating in the Jordan frame.
	
Second, we derive an effective potential directly within the Einstein frame using its Hamilton-Jacobi equation~\eqref{VE}. By applying the approximate relation between the Hubble parameters during slow-roll, $\hat{H}\simeq H/\sqrt{f}$ (derived from Eq.~\eqref{HE} when $\dot{f} \ll Hf$), and the field transformation~\eqref{newp}, we can express this potential, which we denote as $\hat{V}_E$, in terms of the Jordan frame field $\phi$:
	\begin{align}\label{VEE}
		\hat{V}_E \equiv 3\hat{H}^2-2\hat{H}_{,\hat{\phi}}^2\simeq3\left(\frac{H}{\sqrt{1-\alpha}}\right)^2-2\left(\frac{d\hat{H}}{d\phi}\frac{d\phi}{d\hat{\phi}}\right)^2=\frac{3H^2(\phi)\beta-2\mathcal{H}^2(\phi)}{\beta(1-\alpha)},
	\end{align}
where the subscript ``E" denotes the computational path performed in the Einstein frame.
	
A comparison of Eq.~\eqref{VJ} and Eq.~\eqref{VEE} mathematically confirms that $\hat{V}_J \neq \hat{V}_E$. This inequality does not imply a violation of physical principles. Instead, it quantifies the discrepancy between the two approximation schemes. This discrepancy, originating from the non-commutativity of the slow-roll approximation and the conformal transformation, is precisely what we aim to investigate. The differences in these effective potentials will propagate into the inflationary dynamics and reheating predictions, allowing us to quantify the frame-dependence of the Hamilton-Jacobi method itself.

Inflation proceeds when the field $\hat{\phi}$ rolls slowly along $\hat{V}_i$ ($i=E,J$). The slow-roll parameters are defined as
\begin{align}
	\hat{\epsilon} \equiv \frac{1}{2} \left( \frac{1}{\hat{V}_i} \frac{d\hat{V}_i}{d\hat{\phi}} \right)^2, \quad 
	\hat{\eta} \equiv \frac{1}{\hat{V}_i} \frac{d^2\hat{V}_i}{d\hat{\phi}^2}. \label{slowroll}
\end{align}
Inflation ends when either $\hat{\epsilon}_{\rm end} = 1$ or $\hat{\eta}_{\rm end} = 1$, and the number of e-folds is given by
\begin{align}
	\hat{N}_* = \int_{\hat{\phi}_{\rm end}}^{\hat{\phi}_*} \frac{\hat{V}_i}{d\hat{V}_i / d\hat{\phi}} d\hat{\phi}. \label{Nstar}
\end{align}
Here the subscript ``*" represents the moment when the pivot scale crosses outside the Hubble horizon during inflation. 
Inflation predicts the nearly scale-invariant power spectrum of curvature perturbations, with its amplitude given as
\begin{align}\label{As}
	A_{s}=\frac{1}{24\pi^{2}}\frac{\hat{V}_i(\hat{\phi}_{*})}{\hat{\epsilon}(\hat{\phi}_{*})}.
\end{align}
Inflation also gives that the spectral index $\hat{n}_{s}$ of curvature perturbations and the tensor-to-scalar ratio $\hat{r}$ are respectively  
\begin{align}
	\hat{n}_s &\simeq 1 - 6\hat{\epsilon}(\hat{\phi}_*) + 2\hat{\eta}(\hat{\phi}_*), \label{ns} \\
	\hat{r} &\simeq 16 \hat{\epsilon}(\hat{\phi}_*). \label{r}
\end{align}

	\begin{figure}[!ht]
		\centering
		\subfigure[]{\includegraphics[width=0.45\linewidth]{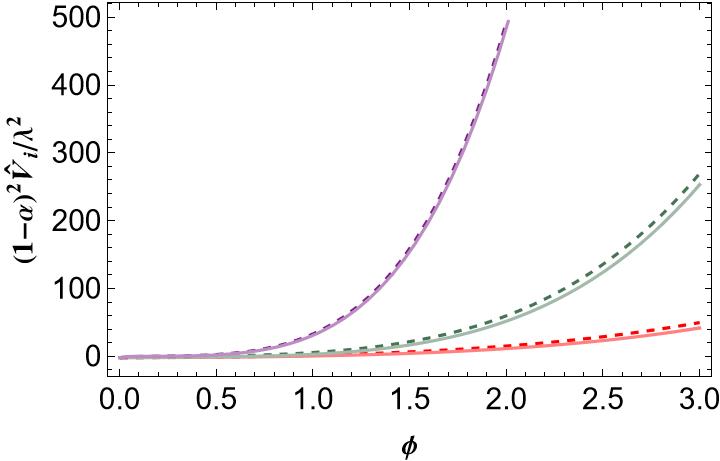}}
		\hspace{0.05\linewidth}
		\subfigure[]{\includegraphics[width=0.45\linewidth]{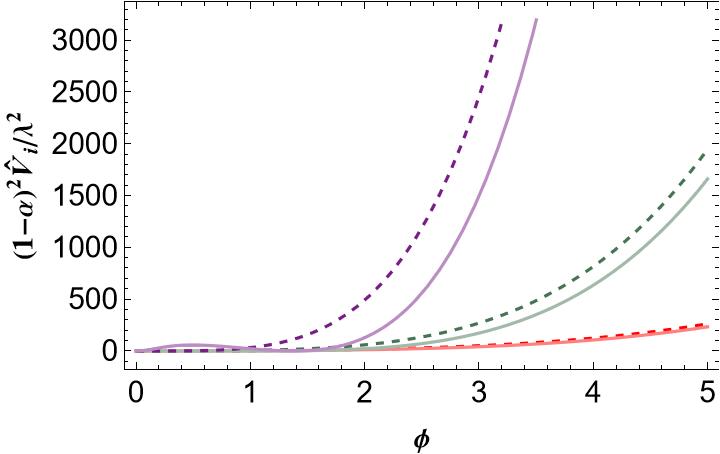}}
		\caption{\label{fig1}
			Effective potentials $(1 - \alpha)^2 \hat{V}_i/\lambda^2$ ($i=J,E$) corresponding to the two computational schemes, in the metric (a) and Palatini (b) formalisms. Solid (dashed) lines correspond to the Jordan-frame scheme (Einstein-frame scheme). The red, green and purple lines correspond to $\xi=-0.1, -1$ and $-10$ for $n=1$.}
	\end{figure}
	
	\begin{figure}[!ht]
		\centering
		\subfigure[]{\includegraphics[width=0.4\linewidth]{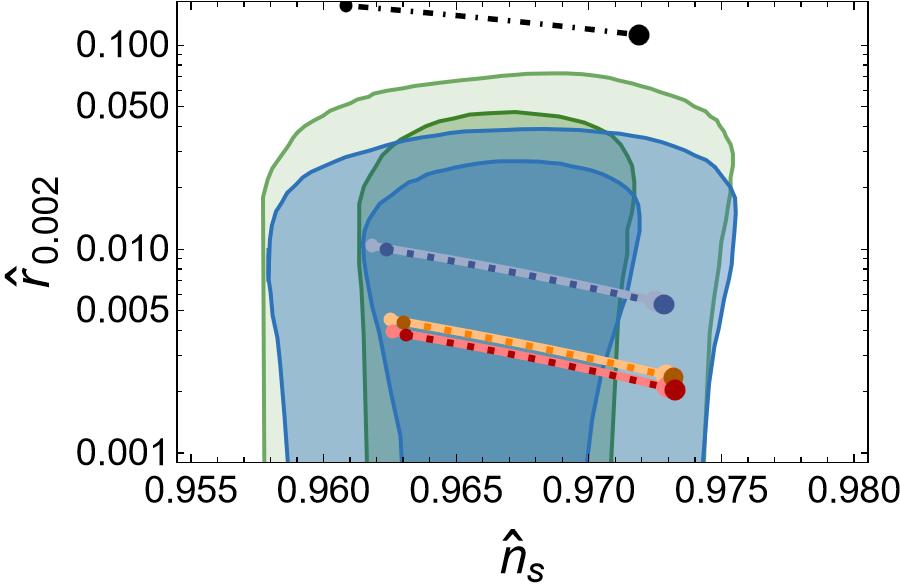}}
		\hspace{0.05\linewidth}
		\subfigure[]{\includegraphics[width=0.4\linewidth]{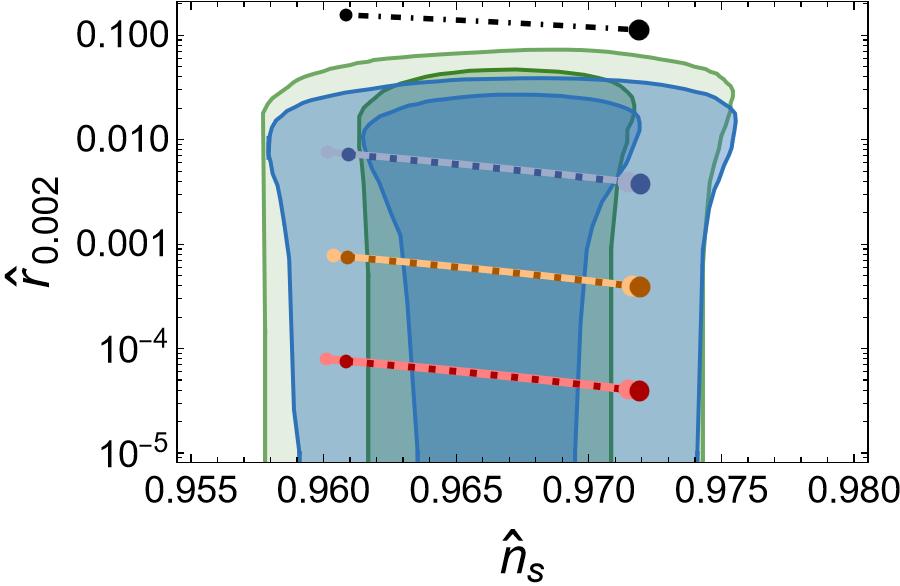}}
		\caption{\label{fig2}
			The predicted values of $\hat{r}$ and $\hat{n}_{\mathrm{s}}$ for the non-minimally coupled inflaton field with various $N_*$ are shown in the metric (a) and Palatini (b) cases.
			The dashed and solid lines correspond to the results derived from the Hamilton-Jacobi method applied in the Einstein and Jordan frames, respectively. The blue, orange, and red lines correspond to $n=1$ with $\xi=-0.1$, $-1$, and $-10$, respectively.
			The black dot-dashed line represents $n=1$ with $\xi=0$, and the small and large dots indicate $\hat{N}_*=50$ and $70$, respectively.
			The green and blue shaded regions depict the constraints on $\hat{n}_{\mathrm{s}}$ and $r$ at the pivot scale $k_*=0.002$ $\mathrm{Mpc}^{-1}$ from the Planck $2018$ CMB observations~\cite{akrami2020planck} and the BICEP/Keck survey~\cite{PhysRevLett.127.151301}, respectively. These regions represent the the $68\%$ and $95\%$ CL contours, shown with dark and light shading, respectively.}
	\end{figure}

To quantify the differences between the two computational schemes, we adopt a power-law form of the Hubble parameter:
	\begin{align}
		H(\phi) = \lambda \phi^n, \label{H1}
	\end{align}
	where $n$ is the model parameter and $\lambda$ is a constant parameter that can be determined by the power spectrum amplitude of  curvature perturbations~\cite{akrami2020planck}:
	\begin{align}
		\ln (10^{10} A_s) = 3.044 \pm 0.014.
	\end{align}
	In Fig.~\ref{fig1}, we present the rescaled effective potentials $(1 - \alpha)^2 \hat{V}_i/\lambda^2$ with respect to the scalar field $\phi$ in the metric (Fig.~\ref{fig1}a) and Palatini (Fig.~\ref{fig1}b) cases. From the figure, we can notice that the two Hamilton-Jacobi approximation schemes yield different potential shapes. As the coupling strength $|\xi|$ increases, the discrepancy between the potentials derived from the Einstein- and Jordan-frame schemes becomes more pronounced in Palatini gravity. This graphical representation provides direct and compelling evidence for the central thesis of our paper: that the choice of computational frame for applying the Hamilton-Jacobi approximation is not trivial and leads to tangible differences in the reconstructed potential, which in turn affect physical observables.

	In Fig.~\ref{fig2}, we show  a comparison between the latest observations and the results of the spectral index $\hat{n}_s$ versus the tensor-to-scalar ratio $\hat{r}$ in the metric (Fig.~\ref{fig2}a) and Palatini (Fig.~\ref{fig2}b) formulations. The inflationary predictions from both computational schemes are in good agreement with the latest Planck CMB data. Notably, the scheme applied in the Einstein frame predicts a slightly larger $\hat{n}_s$ and a smaller $\hat{r}$ compared to the scheme originating from the Jordan frame, reflecting the methodological, frame-dependent differences in the Hamilton-Jacobi formalism.

\section{Reheating Phase in Non-Minimally Coupled Inflation}
\label{sec4}

Having examined the inflationary dynamics and the method-dependent predictions of cosmological observables, we now turn to the post-inflationary reheating phase, which bridges the inflationary epoch and the radiation-dominated era. In this section, we establish the connection between inflationary observables and reheating parameters using the comoving pivot scale observed in the CMB:
\begin{align}
	k_{*} = \hat{a}_* \hat{H}_* = \frac{\hat{a}_*}{\hat{a}_{\mathrm{end}}} \cdot \frac{\hat{a}_{\mathrm{end}}}{\hat{a}_{\mathrm{re}}} \cdot \frac{\hat{a}_{\mathrm{re}}}{\hat{a}_0} \cdot \hat{a}_0 \hat{H}_*,
	\label{kstar}
\end{align}
where the subscripts "$\mathrm{re}$" and "$0$" denote the end of reheating and the present time, respectively. We set $\hat{a}_0 = 1$ for convenience.
The total number of e-folds during reheating is defined as:
\begin{align}
	\hat{N}_{\mathrm{re}} \equiv \ln\left( \frac{\hat{a}_{\mathrm{re}}}{\hat{a}_{\mathrm{end}}} \right).
\end{align}
Thus, the first two factors in Eq.~\eqref{kstar} combine as:
\begin{align}
	\frac{\hat{a}_*}{\hat{a}_{\mathrm{end}}} \cdot \frac{\hat{a}_{\mathrm{end}}}{\hat{a}_{\mathrm{re}}} = e^{-\hat{N}_* - \hat{N}_{\mathrm{re}}}.
	\label{kstar2}
\end{align}
The ratio $\hat{a}_{\mathrm{re}}/\hat{a}_0$ can be related to the reheating temperature $\hat{T}_{\mathrm{re}}$ via entropy conservation in a comoving volume:
\begin{align}
	\hat{g}_{\mathrm{re}} \hat{a}_{\mathrm{re}}^3 \hat{T}_{\mathrm{re}}^3 = \hat{g}_\gamma \hat{T}_\gamma^3 + \frac{7}{8} \hat{g}_\nu \hat{T}_\nu^3 = \left( \frac{43}{11} \right) \hat{T}_\gamma^3,
	\label{entropy}
\end{align}
where $\hat{g}_\mathrm{re}$ is the effective number of relativistic degrees of freedom at the end of reheating (taken to be $106.75$), and $\hat{T}_\gamma = 2.7255$ K is the present CMB temperature. 
We adopt $\hat{g}_\gamma = 2$, $\hat{g}_\nu = 6$, and $\hat{T}_\nu^3 = (4/11)\hat{T}_\gamma^3$ for the calculation. Hence, we arrive at
\begin{align}
	\frac{\hat{a}_{\mathrm{re}}}{\hat{a}_0} = \left( \frac{43}{11 \hat{g}_{\mathrm{re}}} \right)^{1/3} \frac{\hat{T}_\gamma}{\hat{T}_{\mathrm{re}}},
	\label{kstar3}
\end{align}
which, when substituted into Eq.~\eqref{kstar} alongside Eq.~\eqref{kstar2}, gives the reheating temperature:
\begin{align}
	\hat{T}_{\mathrm{re}} = \left( \frac{43}{11\hat{g}_{\mathrm{re}}} \right)^{1/3} \left( \frac{\hat{T}_\gamma}{k_*} \right) \hat{H}_* e^{-\hat{N}_* - \hat{N}_{\mathrm{re}}}.
	\label{Tre}
\end{align}

Once the reheating phase ends, the Universe becomes radiation-dominated. The energy density at this moment is:
\begin{align}
	\hat{\rho}_{\mathrm{re}} = \frac{\pi^2}{30} \hat{g}_{\mathrm{re}} \hat{T}_{\mathrm{re}}^4.
	\label{rhore1}
\end{align}
Assuming a constant effective equation-of-state parameter $\hat{w}_{\mathrm{re}}$ during reheating and applying the continuity equation, the energy density evolves as:
\begin{align}
	\hat{\rho}_{\mathrm{re}} = \hat{\rho}_{\mathrm{end}} e^{-3 \hat{N}_{\mathrm{re}} (1 + \hat{w}_{\mathrm{re}})},
	\label{rhore2}
\end{align}
where $\hat{\rho}_{\mathrm{end}}$ is the energy density at the end of inflation, approximated as:
\begin{align}
	\hat{\rho}_{\mathrm{end}} = \frac{3}{2} \hat{V}_i(\hat{\phi}_{\mathrm{end}}).
	\label{rhoend}
\end{align}
Combining Eqs.~\eqref{Tre}–\eqref{rhoend}, we derive a expression relating inflationary parameters to reheating quantities:
\begin{align}
	(3\hat{w}_{\mathrm{re}} - 1) \hat{N}_{\mathrm{re}} = \ln \left( \frac{45}{\pi^2 \hat{g}_{\mathrm{re}}} \right) + \ln \hat{V}_i(\hat{\phi}_{\mathrm{end}}) + \frac{4}{3} \ln \left( \frac{11 \hat{g}_{\mathrm{re}}}{43} \right) 
	 + 4\ln \left( \frac{k_*}{\hat{H}_* \hat{T}_\gamma} \right) + 4\hat{N}_*.
	\label{reheating_equation}
\end{align}
This equation is pivotal in constraining the inflationary model based on CMB observations. Notably, it becomes independent of reheating details when either $\hat{w}_{\mathrm{re}} = 1/3$, corresponding to an equation-of-state matching radiation, or $\hat{N}_{\mathrm{re}} = 0$, corresponding to instantaneous reheating.

\section{Numerical Results and Comparative Analysis}
\label{sec5}
	In this section, we present a detailed numerical investigation of the reheating phase in non-minimally coupled inflationary models, utilizing the two Hamilton-Jacobi computational schemes outlined in Sec. III C.
We analyze the predictions of the metric and Palatini formulations  and compare the results  with current observational constraints.
From Eq.~\eqref{ns}, the scalar spectral index $\hat{n}_s$ is determined by  the field value $\hat{\phi}_*$,  the model parameters $\xi$ and $n$. 
Since $\hat{N}_*$ can also be expressed in terms of $\hat{\phi}_*$, it follows from Eqs.~\eqref{reheating_equation} and \eqref{Tre} that $\hat{N}_{\mathrm{re}}$ and $\hat{T}_{\mathrm{re}}$ can be consistently written as functions of $\hat{n}_s$ for fixed $\hat{w}_{\mathrm{re}}$, $\xi$, and $n$.

\subsection{Metric Formalism}

\begin{figure}[t]
	\centering  
	\subfigtopskip=2pt 
	\subfigbottomskip=2pt 
	\subfigcapskip=-5pt 
	\subfigure{
		\includegraphics[width=0.31\linewidth]{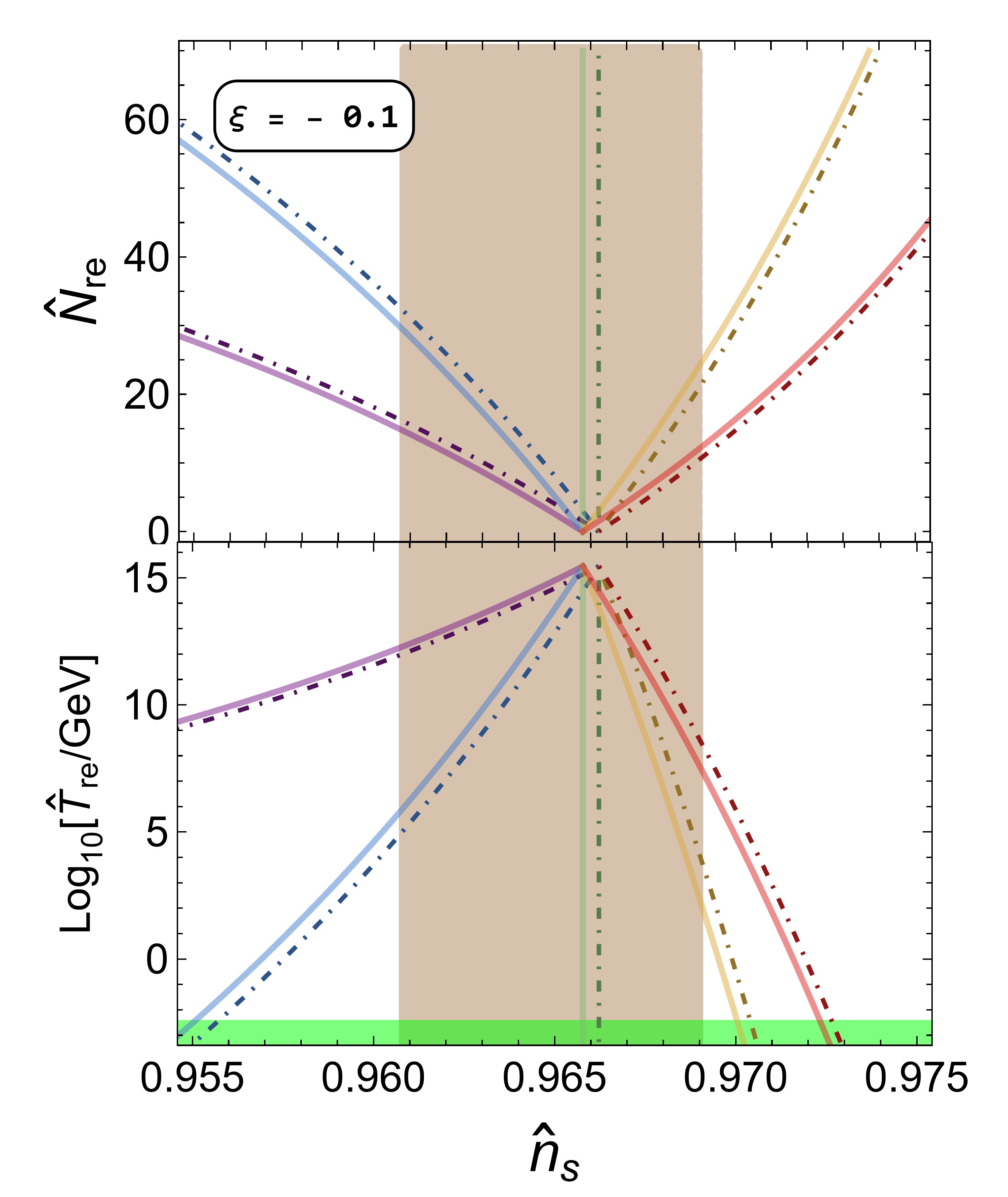}
	}
	\subfigure{
		\includegraphics[width=0.31\linewidth]{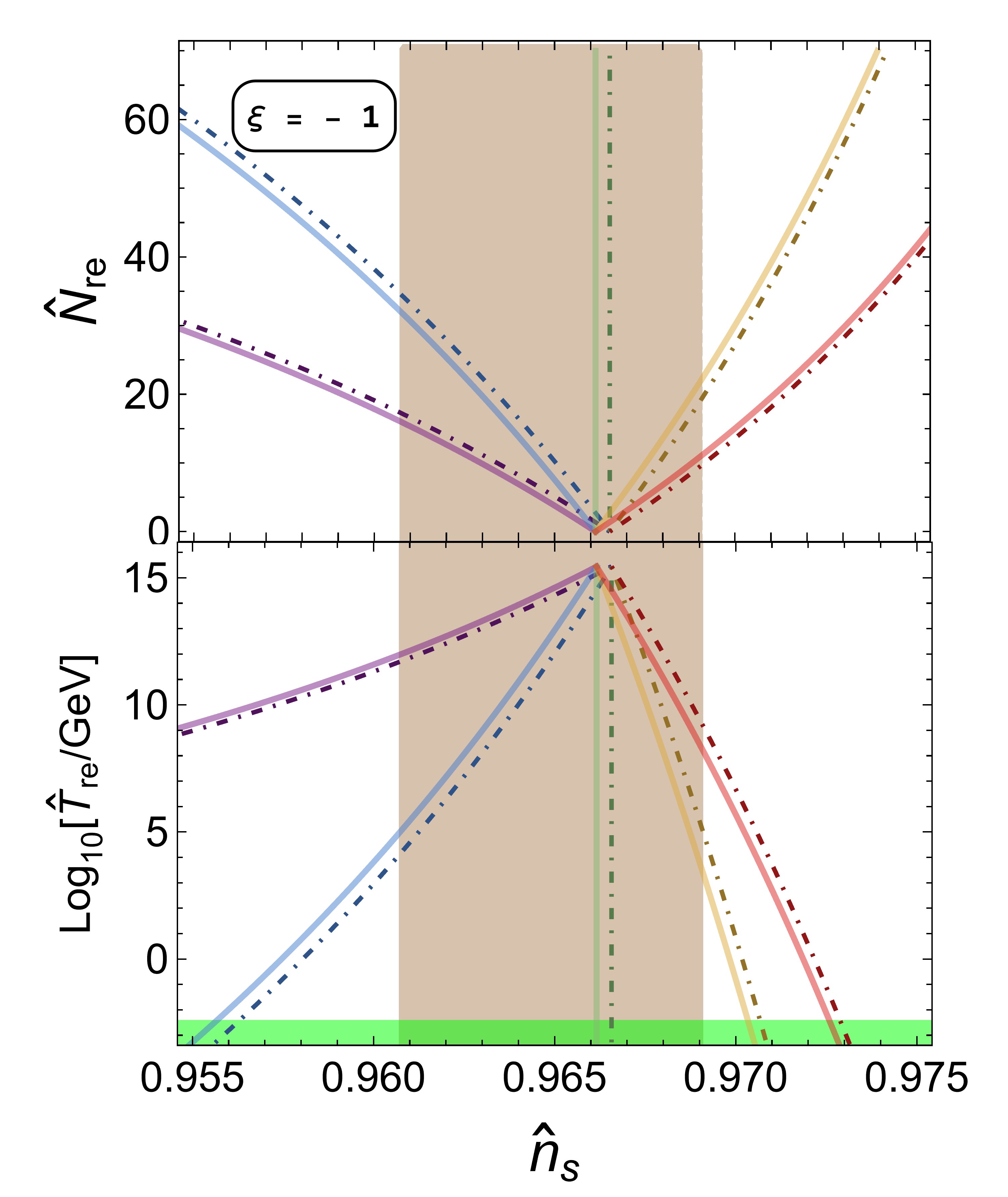}
	}
    \subfigure{
		\includegraphics[width=0.31\linewidth]{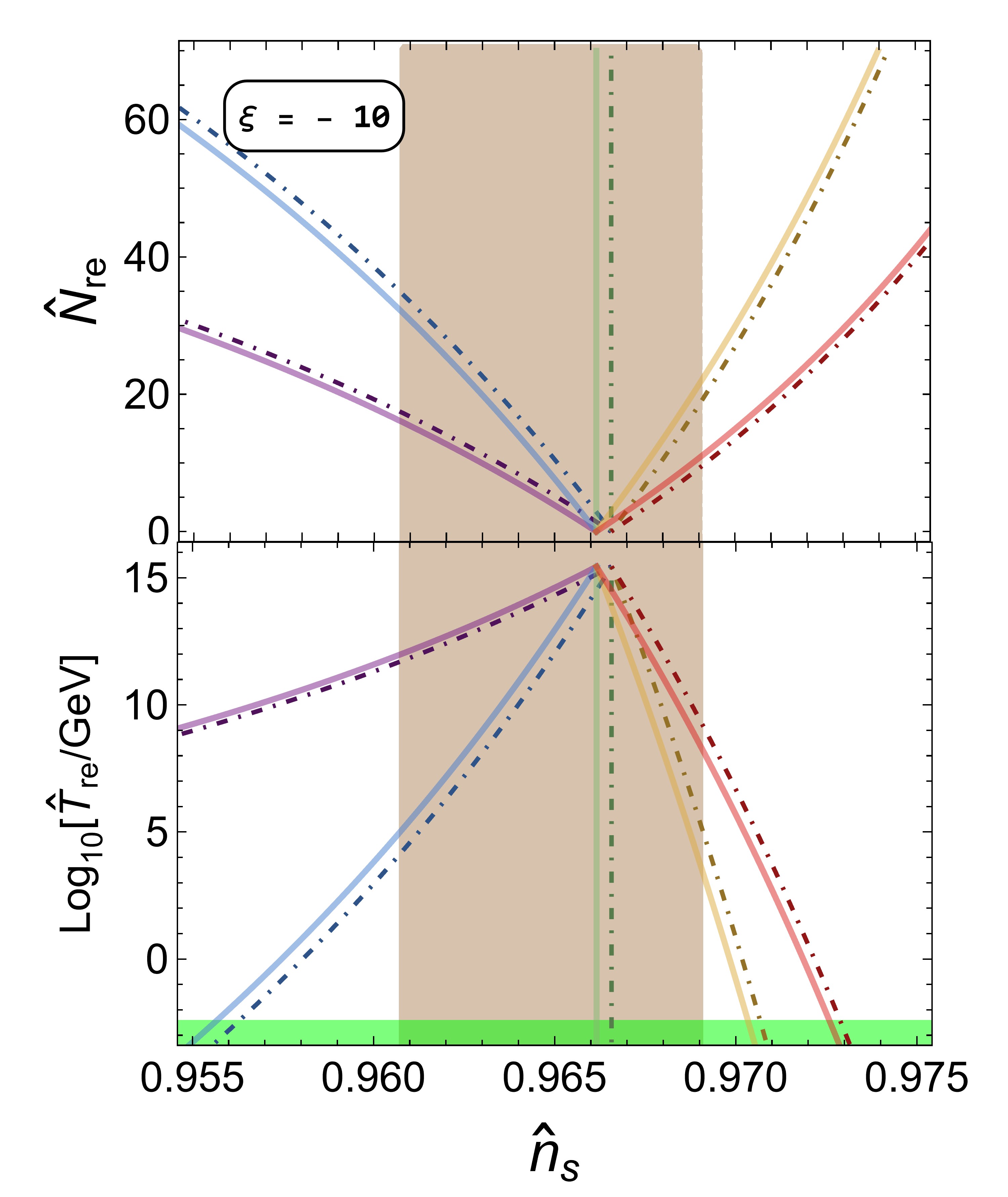}
	}
	\caption{\label{fig3} Reheating e-folding number $\hat{N}_{\mathrm{re}}$ and reheating temperature $\log_{10}[\hat{T}_{\mathrm{re}}/\mathrm{GeV}]$ as functions of $\hat{n}_s$ for $n=1$ with varying $\xi$ in the metric formulation. Solid and dot-dashed lines represent the Jordan-frame scheme and Einstein-frame scheme results, respectively. The purple, blue, green, orange and red curves correspond to $w_{\mathrm{re}}= -1/3, 0, 1/3, 2/3$ and $1$, respectively.
    The brown band indicates the $1\sigma$ bound on $\hat{n}_s$ from Planck $2018$ TT, TE, EE+lowE+lensing~\cite{akrami2020planck}. 
    Temperatures below $T_\mathrm{re} < 4$MeV shown as the green region are ruled out by the BBN~\cite{PhysRevLett.82.4168,PhysRevD.62.023506, PhysRevD.70.043506, Hasegawa_2019}.
}	
\end{figure}

	As shown in Fig.~\ref{fig3}, in the metric case, the reheating e-folding number $\hat{N}_{\mathrm{re}}$ and the reheating temperature $\log_{10}[\hat{T}_{\mathrm{re}}/\mathrm{GeV}]$ are presented as functions of $\hat{n}_s$ for $n=1$ and several values of $\xi$. 
	The colored curves correspond to different choices of the reheating equation-of-state parameter $\hat{w}_{\mathrm{re}} \in \{-1/3, 0, 1/3, 2/3, 1\}$, where $\hat{w}_{\mathrm{re}} = -1/3$ approximates the effective value near the end of inflation and the upper limit $\hat{w}_{\mathrm{re}} \leq 1$ ensures causal consistency. 
	Clearly, all curves shift toward larger $\hat{n}_s$ values as $|\xi|$ increases, and the Hamilton-Jacobi scheme applied in the Einstein frame yields a slightly larger $\hat{n}_s$ than the scheme applied in the Jordan frame for the same set of parameters. This behavior is consistent with the inflationary predictions presented in earlier sections. This discrepancy reflects the intrinsic difference between the two approximation schemes.
	In addition, instantaneous reheating is allowed by the Planck CMB observations for all values of $\xi$.

\begin{figure}[t]
	\centering  
	\subfigure{
		\includegraphics[width=0.31\linewidth]{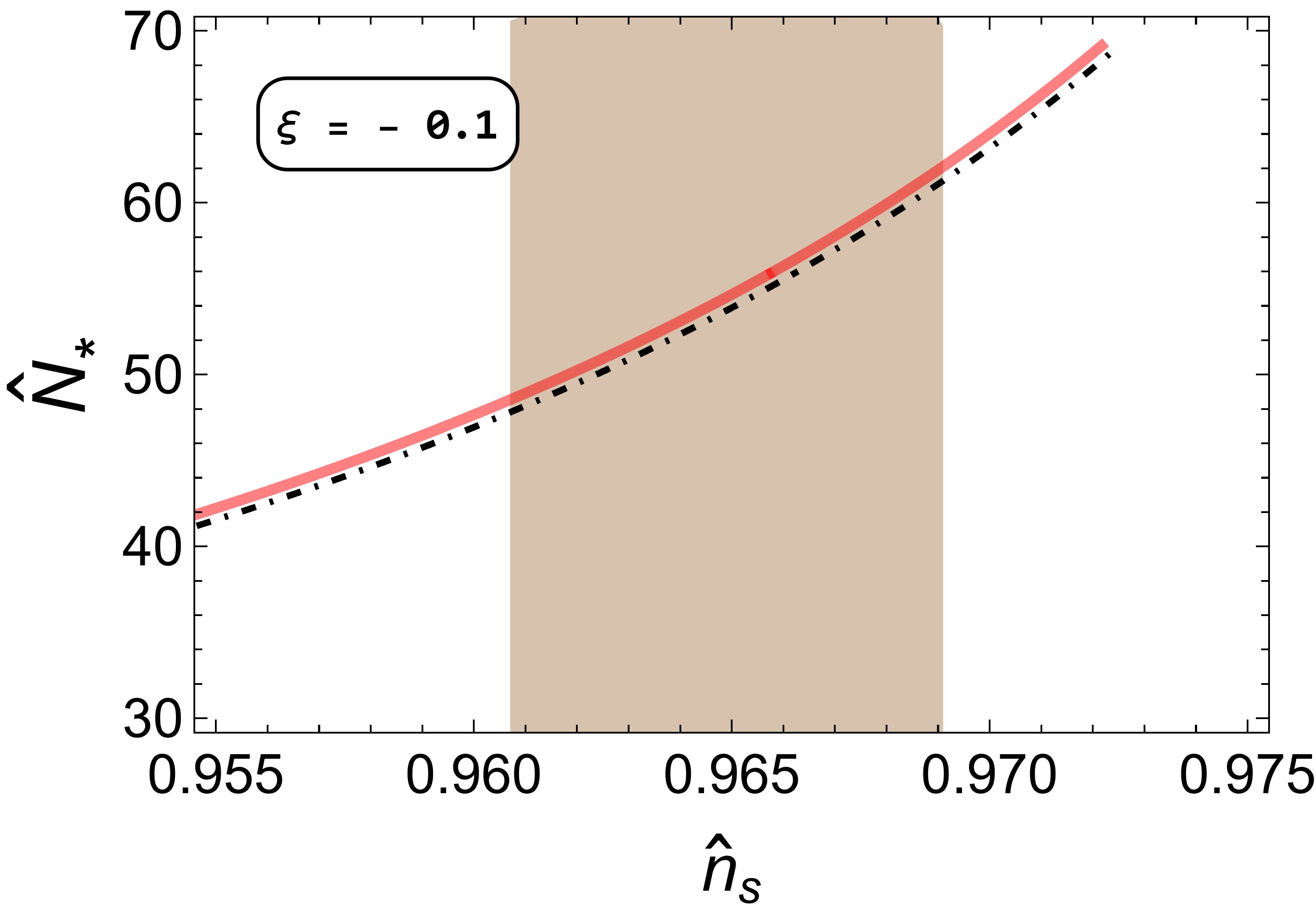}
	}
	\subfigure{
		\includegraphics[width=0.31\linewidth]{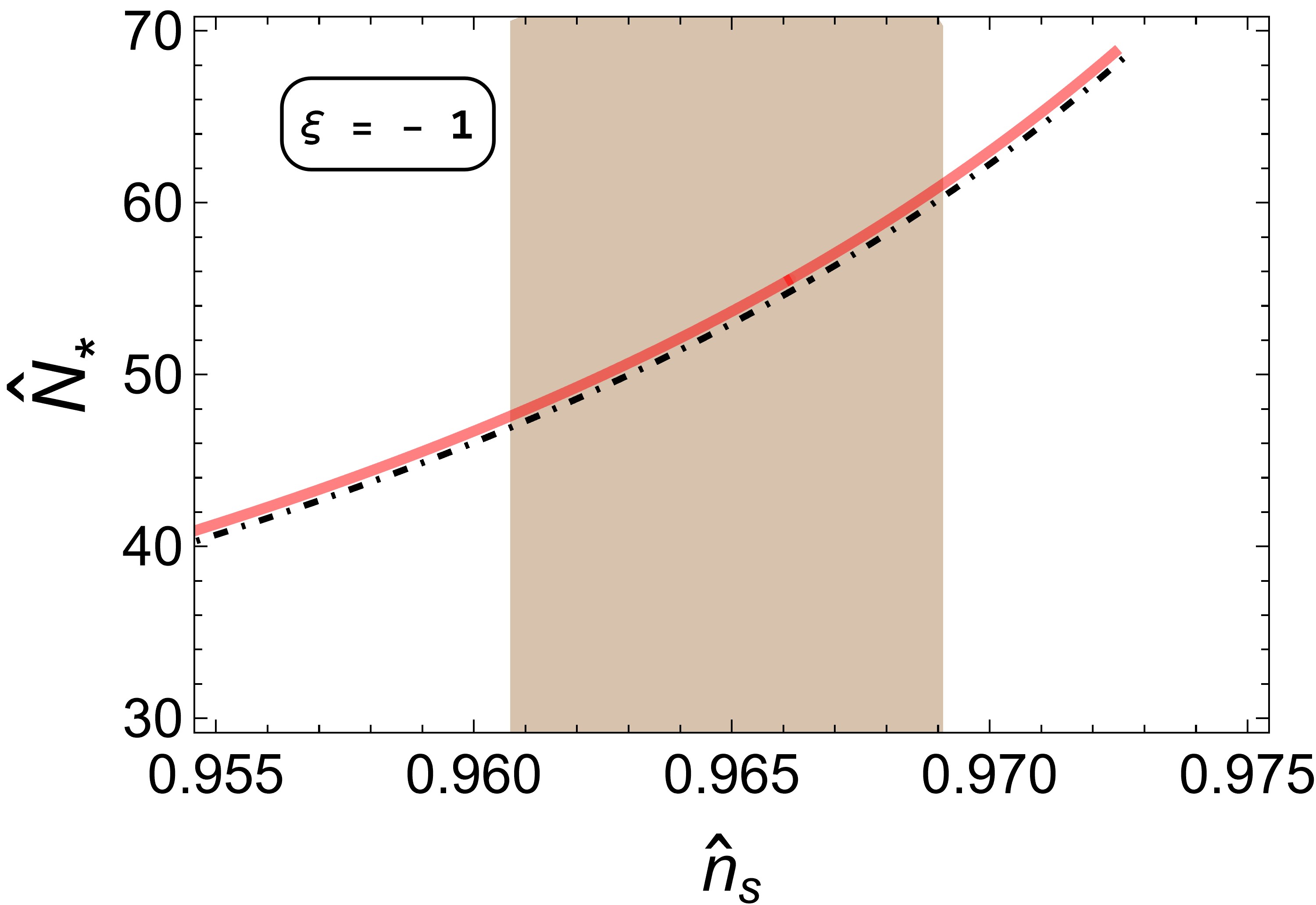}
	}
    \subfigure{
		\includegraphics[width=0.31\linewidth]{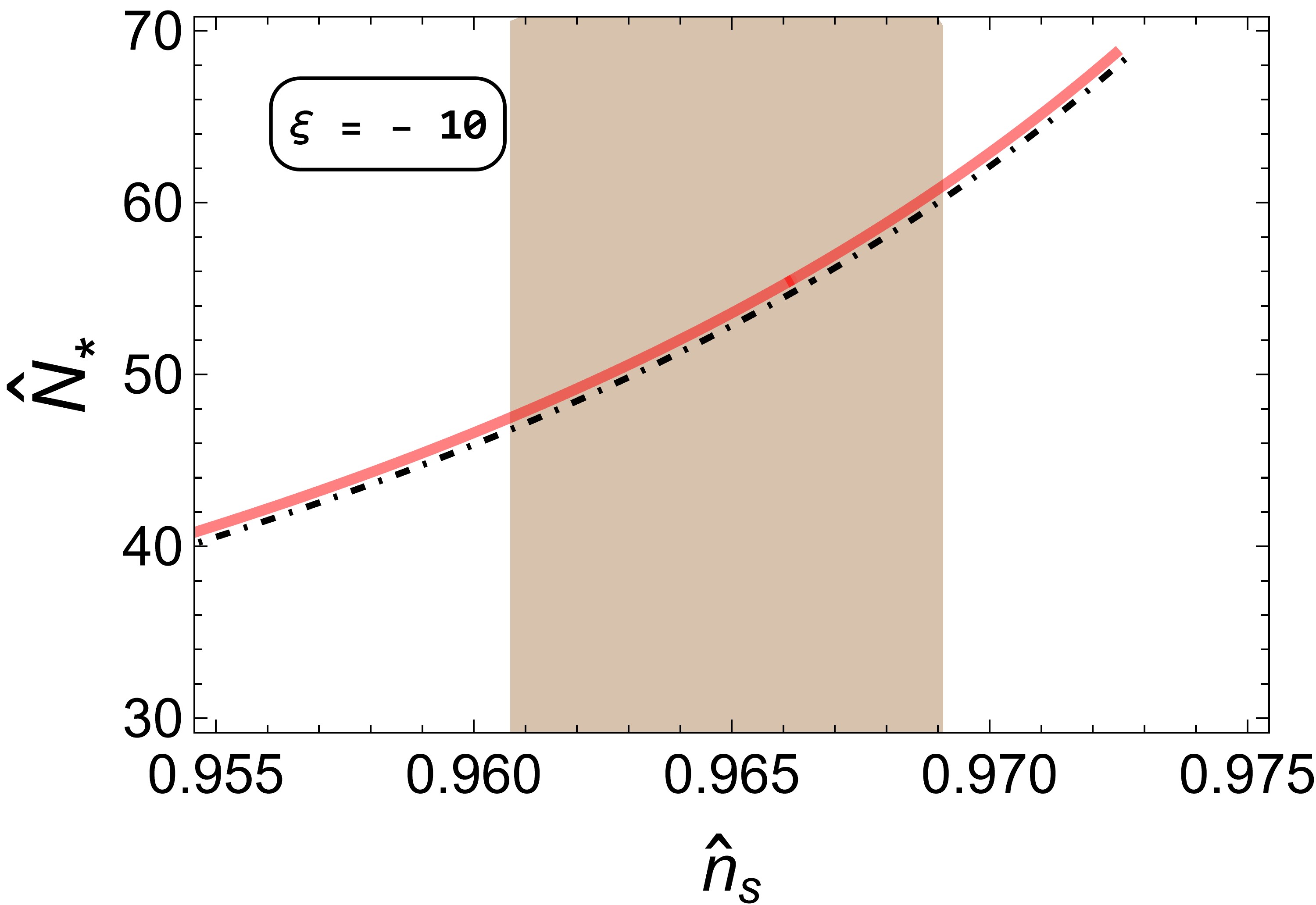}
	}
	\caption{\label{fig4} Predicted $\hat{N}_*$ versus $\hat{n}_s$ for $n=1$ with varying $\xi$ in the metric formalism. Solid and dashed curves denote Jordan- and Einstein-frame scheme results, respectively.
     The brown region  indicates the $1\sigma$ bound on $\hat{n}_s$ from the Planck $2018$ TT, TE, EE+lowE+lensing~\cite{akrami2020planck}. 
    }
\end{figure}

	In Fig.~\ref{fig4}, we plot the predicted values of $\hat{N}_*$ as functions of $\hat{n}_s$ for various values of  $\xi$, and the corresponding allowed ranges are summarized in Table~\ref{TABLE 1}. We find that larger $|\xi|$ leads to a smaller $\hat{N}_*$. 
	Additionally, there is a slight discrepancy between the predictions of the Einstein and Jordan frame schemes: for each value of $\xi$, the Jordan frame scheme consistently predicts slightly larger $\hat{N}_*$ values, typically within $1$ e-fold.
	Overall, the predicted $\hat{N}_*$ values in the metric case typically lie in the range $47 \lesssim \hat{N}_* \lesssim 62$, depending on the specific choice of $\xi$ and the computational scheme considered.

	\begin{table*}[t]
		\renewcommand{\arraystretch}{1.5}
		\setlength{\tabcolsep}{8pt}
		\caption{\label{TABLE 1} Predicted range of $\hat{N}_*$ consistent with Planck 2018 CMB observations in the metric case for different values of $\xi$.}
		\centering
		\begin{tabular}{c c c c}
			\hline\hline
			Scheme Origin & $\xi = -0.1$ & $\xi = -1$ & $\xi = -10$ \\
			\hline
			Jordan & $48.5 \lesssim \hat{N}_* \lesssim 62.1$ & $47.6 \lesssim \hat{N}_* \lesssim 61.1$ & $47.5 \lesssim \hat{N}_* \lesssim 61.0$ \\
			Einstein & $47.8 \lesssim \hat{N}_* \lesssim 61.3$ & $46.9 \lesssim \hat{N}_* \lesssim 60.3$ & $46.8 \lesssim \hat{N}_* \lesssim 60.2$ \\
			\hline\hline
		\end{tabular}
	\end{table*}

\subsection{Palatini Formalism}

\begin{figure}[t]
	\centering  
	\subfigure{
		\includegraphics[width=0.31\linewidth]{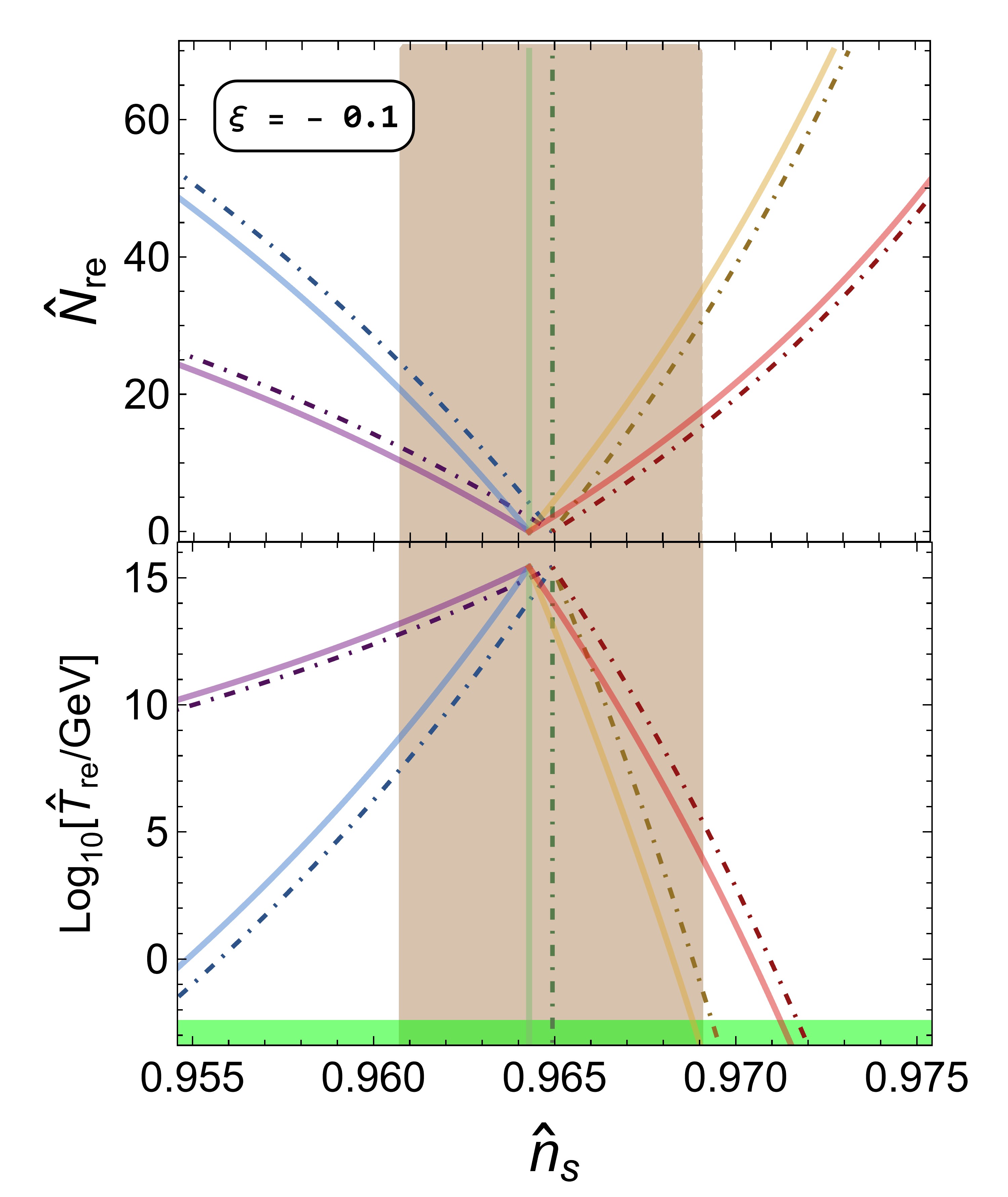}
	}
	\subfigure{
		\includegraphics[width=0.31\linewidth]{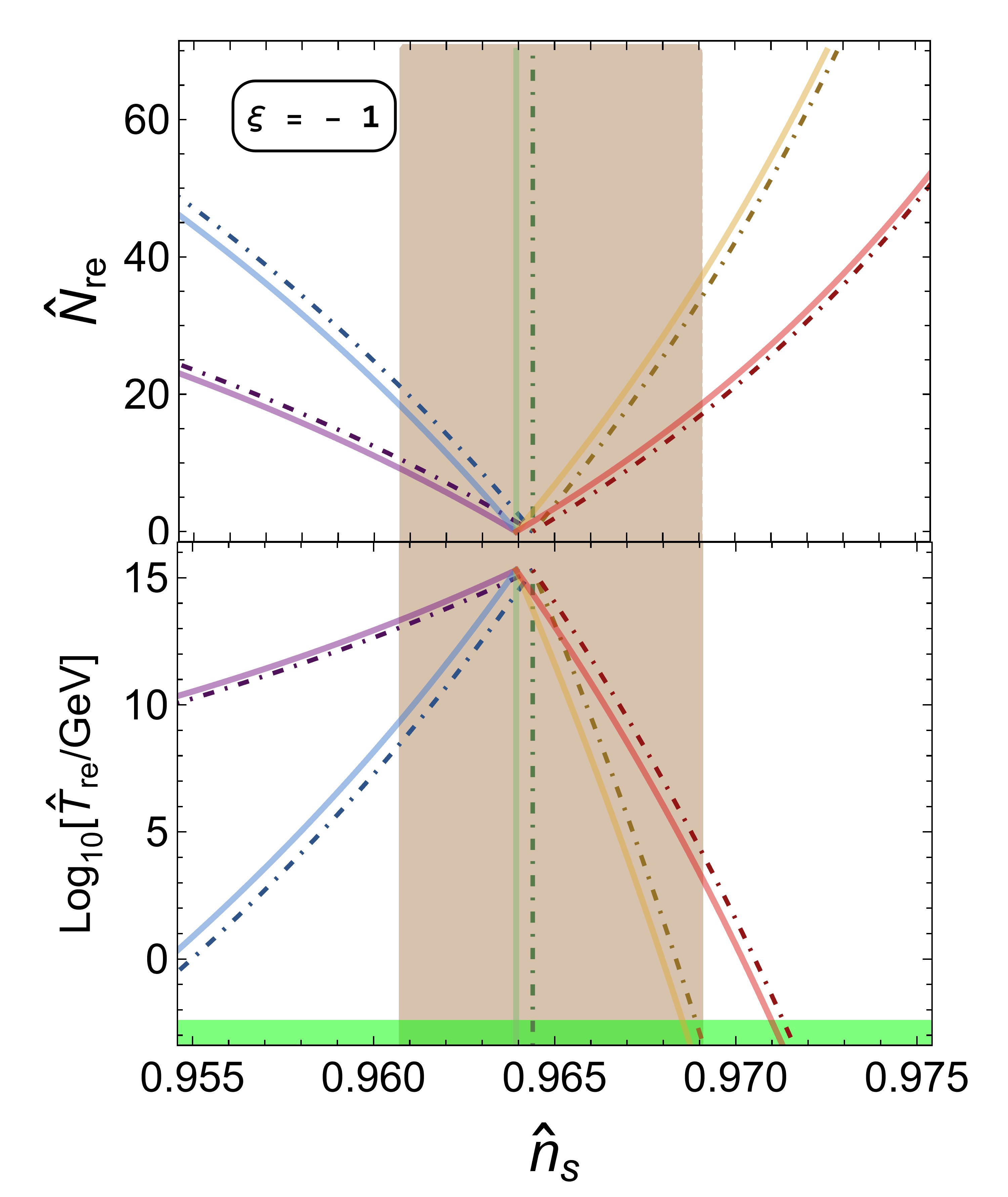}
	}
    \subfigure{
		\includegraphics[width=0.31\linewidth]{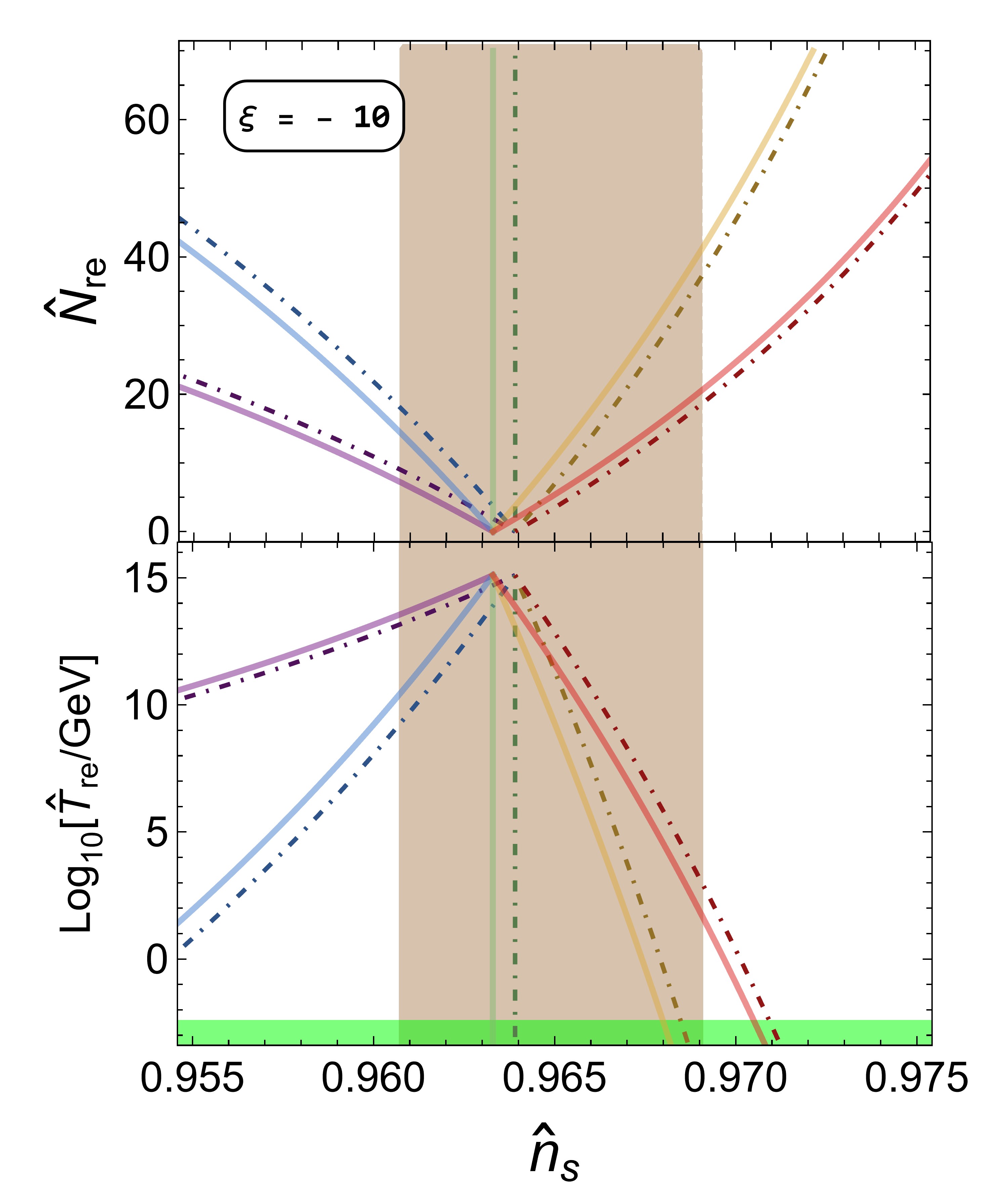}
	}
	\caption{\label{fig5} Reheating predictions in the Palatini case for $n=1$: $\hat{N}_{\mathrm{re}}$ and $\log_{10}[\hat{T}_{\mathrm{re}}/\mathrm{GeV}]$ as functions of $\hat{n}_s$ for varying $\xi$. Line styles and shaded regions follow the conventions of Fig.~\ref{fig3}.}
\end{figure}

We now turn to the Palatini formulation. Fig.~\ref{fig5} shows  $\hat{N}_{\mathrm{re}}$ and $\log_{10}[\hat{T}_{\mathrm{re}}/\mathrm{GeV}]$ versus $\hat{n}_s$ for different $\xi$. Compared to the metric case, the curves now shift to the left as $|\xi|$ increases, reflecting the distinct dynamics induced by the Palatini connection structure. Instantaneous reheating remains allowed for all $\xi$ values, as indicated by the consistent intersection of the curves with the Planck $1\sigma$ region.

\begin{figure}[t]
	\centering  
	\subfigure{
		\includegraphics[width=0.31\linewidth]{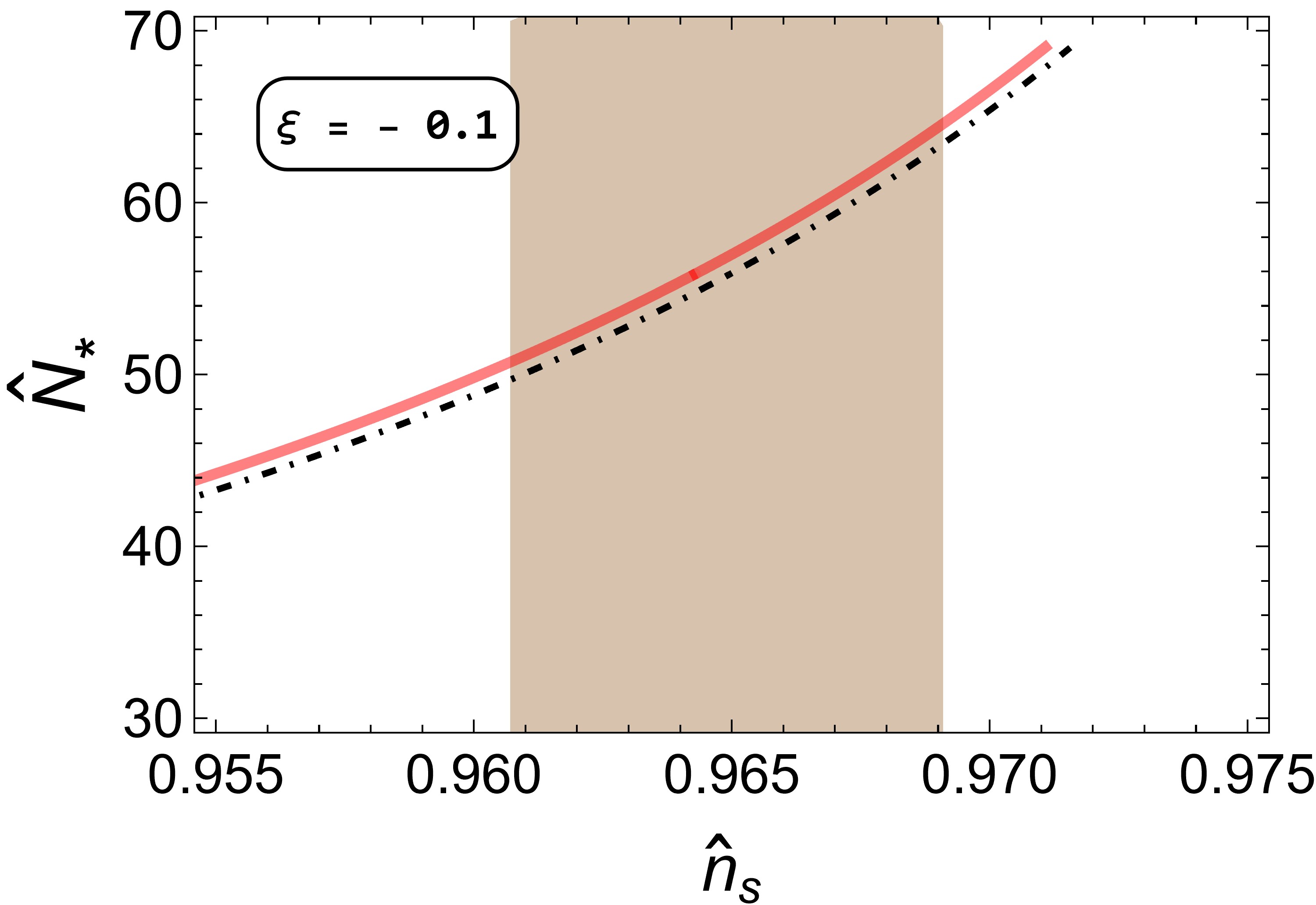}
	}
    \subfigure{
		\includegraphics[width=0.31\linewidth]{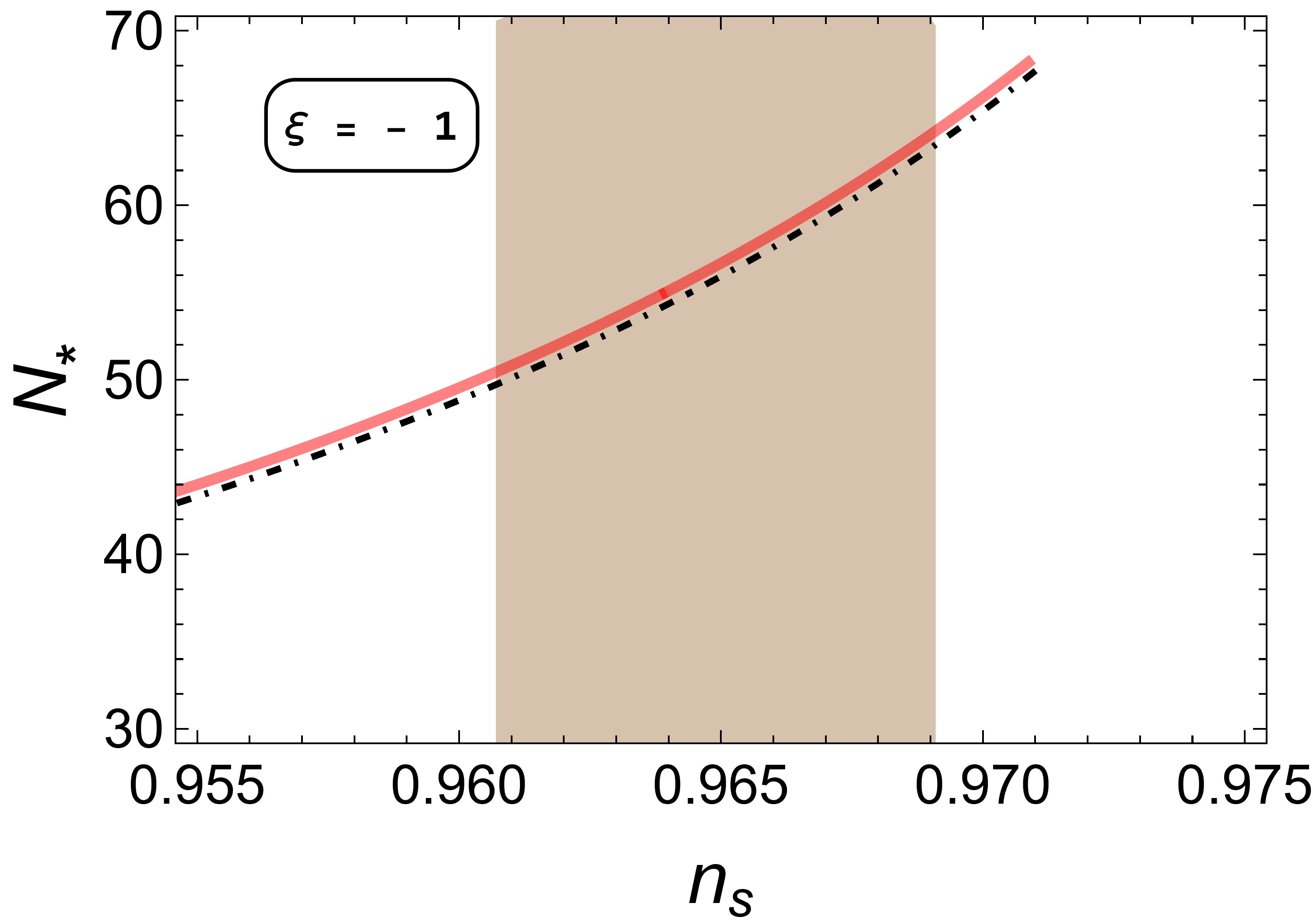}
	}
    \subfigure{
		\includegraphics[width=0.31\linewidth]{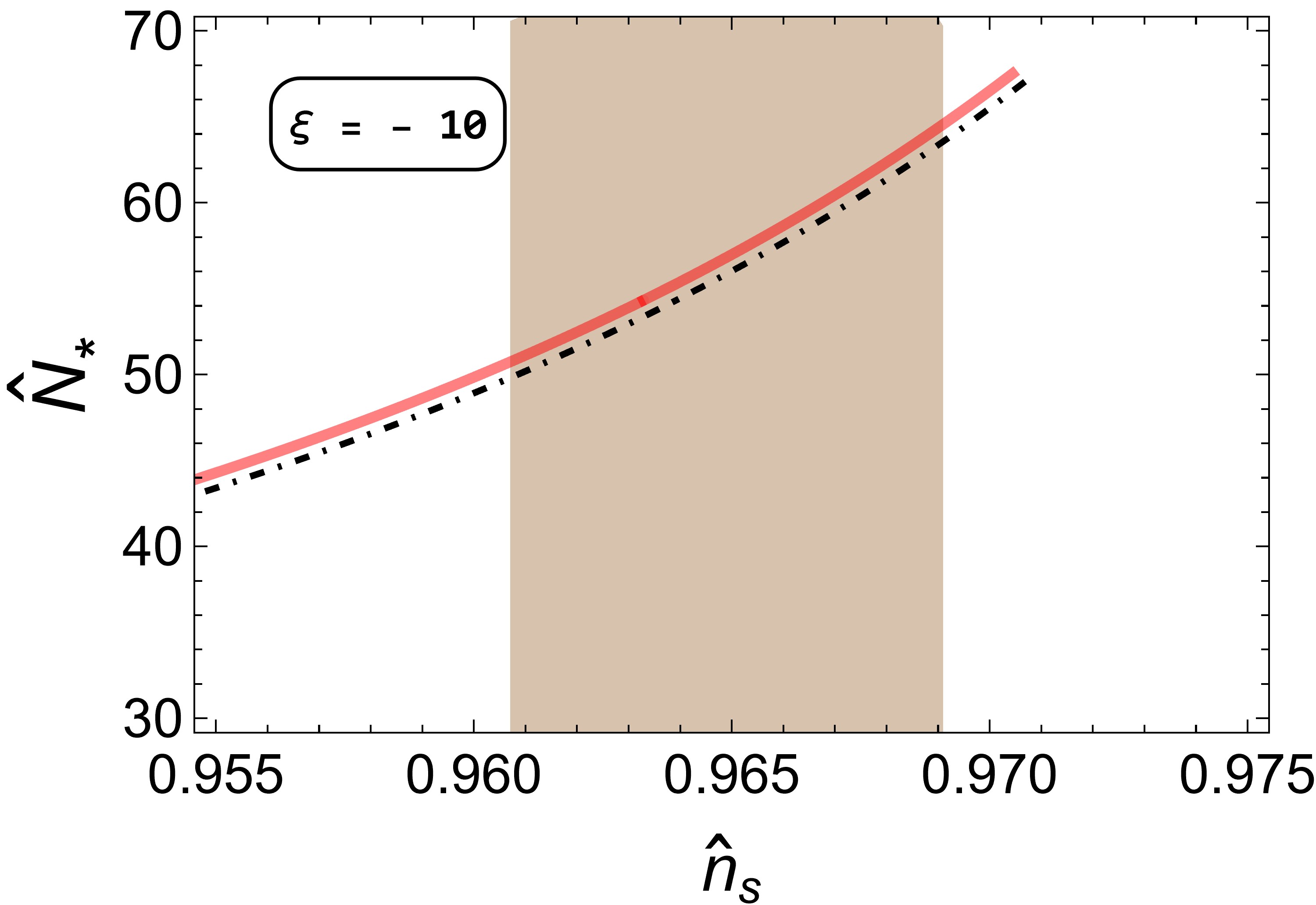}
	}
	\caption{\label{fig6} Evolution of $\hat{N}_*$ versus $\hat{n}_s$ for $n=1$ and varying $\xi$ in the Palatini formulation. Line styles and color coding follow Fig.~\ref{fig4}.}
\end{figure}

	In Fig.~\ref{fig6}, we plot the evolution of $\hat{N}_*$ as a function of $\hat{n}_s$ for different values of $\xi$ in the Palatini formulation, and the allowed ranges of $\hat{N}_*$ are summarized in Table~\ref{TABLE 2}. Unlike the metric case, the variation in $\hat{N}_*$ with respect to $\xi$ is negligible, and the overall values of $\hat{N}_*$ are slightly higher.
	Furthermore, the difference between the Einstein and Jordan frame scheme predictions is slightly more pronounced in Palatini gravity.
	For instance, the Jordan frame scheme consistently yields $\hat{N}_*$ values about $1$ e-fold larger than those in the Einstein frame scheme, highlighting the enhanced methodological frame dependence of the reheating predictions in this formulation.

\begin{table*}[t] 
    \renewcommand{\arraystretch}{1.5} 
    \setlength{\tabcolsep}{8pt}
    \caption{\label{TABLE 2} Allowed range of $\hat{N}_*$ consistent with Planck CMB constraints in the Palatini formulation for various values of $\xi$.}
    \centering 
    \begin{tabular}{c c c c}
        \hline\hline
        Scheme Origin & $\xi = -0.1$ & $\xi = -1$ & $\xi = -10$ \\
        \hline
        Jordan   & $50.7 \lesssim \hat{N}_* \lesssim 64.5$ & $50.5 \lesssim \hat{N}_* \lesssim 64.3$ & $50.7 \lesssim \hat{N}_* \lesssim 64.6$ \\
        Einstein & $49.6 \lesssim \hat{N}_* \lesssim 63.4$ & $49.6 \lesssim \hat{N}_* \lesssim 63.4$ & $49.8 \lesssim \hat{N}_* \lesssim 63.6$ \\
        \hline\hline
    \end{tabular}
\end{table*}

	Overall, while both gravitational formalisms exhibit method-dependent features, the Palatini case yields more stable $\hat{N}_*$ predictions against variations in $\xi$, and a clearer distinction between the Einstein and Jordan frame schemes. The predicted values in this case typically fall within the range $49 \lesssim \hat{N}_* \lesssim 65$, depending on the scheme choice.

\section{Conclusions}\label{sec6}

	The Hamilton-Jacobi formalism offers a direct and insightful approach to describing inflationary dynamics, particularly when exploring extended theories of gravity beyond GR. In this work, we have systematically studied non-minimally coupled inflationary models characterized by the coupling function $f(\phi) = 1 - \xi \phi^2$, quantifying the methodological impact of applying the Hamilton-Jacobi approximation in either the Jordan or the Einstein frame. We compared the results under both the metric and Palatini formalisms.
	We computed inflationary observables such as the scalar spectral index $\hat{n}_s$ and the tensor-to-scalar ratio $\hat{r}$, and compared them with the latest Planck $2018$ data. All scenarios investigated yield predictions consistent with observational bounds. However, quantifiable differences arise depending on the computational scheme chosen. In particular, applying the Hamilton-Jacobi approximation in the Einstein frame typically predicts a slightly larger $\hat{n}_s$ and a smaller $\hat{r}$ compared to applying it in the Jordan frame first, with these deviations being more pronounced in the Palatini formulation.
	
	We further extended our analysis to the reheating epoch, establishing a connection between inflationary dynamics and post-inflationary observables. By scanning over a wide range of equation-of-state parameters $\hat{w}_{\mathrm{re}}$ and non-minimal coupling values $\xi$, we explored how reheating predictions evolve under different assumptions. We find that instantaneous reheating remains viable in all scenarios. The discrepancies between the two computational schemes persist and are amplified in the reheating analysis. In the metric formalism, increasing $|\xi|$ leads to a rightward shift in the predictions in the $\hat{n}_s$–$\hat{N}_{\mathrm{re}}$ plane, while the Palatini formulation exhibits a leftward shift.
	Additionally, the difference between the Einstein and Jordan scheme predictions for $\hat{N}_*$ is found to be more significant in the Palatini formulation, underscoring its heightened sensitivity to the choice of computational frame.
	
In summary, our study highlights that the choice of conformal frame in which an approximation scheme like the Hamilton-Jacobi formalism is applied can significantly impact both inflationary and reheating predictions in non-minimally coupled inflation. The Hamilton-Jacobi method serves as a powerful tool to trace these differences, providing a novel probe into the robustness of theoretical predictions and the subtle interplay between approximations and frame transformations in modified gravity.

\begin{acknowledgments}
    This work was supported by the National Natural Science Foundation of China (Grant No. 12505075), the Chengdu Normal University Talent Introduction Scientific Research Special Project (Grant No. YJRC202443), and the Research Incentive Program for Doctors Joining Shanxi (Grant No. Z20240219).
\end{acknowledgments}


%

\end{document}